\documentclass[11pt, a4paper,english,superscriptaddress,aps]{revtex4}
\usepackage{amsmath,amssymb,graphicx}
\makeatletter
\usepackage{babel}
\usepackage[active]{srcltx}
\usepackage{graphicx,color}
\usepackage{changebar}
\usepackage{hyperref}
\newcommand{\be}{\begin{equation}}
\newcommand{\ee}{\end{equation}}
\newcommand{\bea}{\begin{eqnarray}}
\newcommand{\eea}{\end{eqnarray}}

\begin{document}

\title{G\"{o}del-type solutions within the $f(R,Q)$ gravity}

\author{F. S. Gama, J. R. Nascimento, A. Yu. Petrov, P. J. Porf\'{i}rio}
\affiliation{Departamento de F\'{\i}sica, Universidade Federal da Para\'{\i}ba\\
 Caixa Postal 5008, 58051-970, Jo\~ao Pessoa, Para\'{\i}ba, Brazil}
\email{fisicofabricio@yahoo.com.br, jroberto,petrov@fisica.ufpb.br, pporfirio89@gmail.com}

\author{A. F. Santos}
\affiliation{Instituto de F\'{\i}sica, Universidade Federal do Mato Grosso, \\
78060-900, Cuiab\'{a}, Mato Grosso, Brazil}

\email{alesandroferreira@fisica.ufmt.br}

\begin{abstract}
 In this paper, we deal with the $f(R,Q)$ gravity whose action depends, besides of the scalar curvature $R$, on the higher-derivative invariant $Q=R_{\mu\nu}R^{\mu\nu}$. In order to compare this theory with the usual General Relativity (GR), we verify the
consistency of G\"{o}del-type solutions within the $f(R,Q)$ gravity and discuss the related causality issues. Explicitly, we show that in the $f(R,Q)$ gravity there are new G\"{o}del-type completely causal solutions having no analogue in the general relativity. In particular, a remarkable G\"{o}del-type solution corresponding to the conformally flat space and maximally symmetric for physically well-motivated matter sources, with no necessity of cosmological constant, has been considered.  We demonstrate that, in contrast to GR framework, $f(R,Q)$ gravity supports new vacuum solutions with the requirement for the cosmological constant to be non-zero. Finally, causal solutions are obtained for a particular choice $f(R,Q)=R+\alpha R^2+ \beta Q$.
\end{abstract}

\maketitle

\section{Introduction}

 The GR is known to be the successful theory of gravity, its predictions are in accordance with tests realized in solar system, the so-called classical tests, for example, the precession of the perihelion of Mercury, as well as with the recent detection of gravitational waves \cite{LIGO1, LIGO2, LIGO3}. Nonetheless, it turns out that the Einstein gravity fails in some aspects, which leads to interest to search for its possible consistent generalizations. Basically, there are two main problems having no solution within the framework of the GR: the first one takes place on a phenomenological perspective that arises as one of the most enigmatic problems in physics, the accelerated expansion of the Universe. It is confirmed by observational data from Type Ia supernovae \cite{Riess, Riess2, Perlmutter}, from cosmic microwave background (CMB) measurements \cite{boomerang, boomerang2, WMAP} and studies of large structures \cite{LAnderson, Samushia}. The second reason, purely theoretical, is related to issues on quantization of gravity, since, as it is well known, the Einstein gravity is a non-renormalizable theory \cite{Veltman}. Therefore, in quantum regime the GR does not present a consistent quantum gravity theory.  It is expected that in this regime new degrees of freedom become important.

A possible way out for these issues is based on introducing the modifications of Einstein gravity. There are various modified gravity theories involving new fields, the most known ones are scalar-tensor theories involving a coupling of a non-matter scalar field to gravity. On the other hand, it is also possible to introduce new degrees of freedom by considering model involving higher curvature invariants such as $R^2$ and $R_{\mu\nu}R^{\mu\nu}$ in addition to Einstein-Hilbert action. It has been shown in \cite{Stelle} that in particular case where the model is composed by all quadratic curvature invariants added to Einstein-Hilbert action, one gets a renormalizable theory. However, for the renormalizability one pays the price of introducing ghost-like modes. Furthermore, higher curvature terms come up in others approach, for example, they naturally appear in string theory after dimensional reduction process \cite{Polchinski}. It should be noted that even the above-mentioned results of observations of gravitational waves do not rule out completely the possibility of alternative gravity models, see the discussion in \cite{Konoplya:2016pmh}.

 Despite the fact that ghost-like modes are present, higher curvature theories have been receiving a recent overwhelming interest from the aforementioned fact that these theories are renormalizable. To verify the consistency of these theories with the cosmological observations, it is interesting to examine the behavior of the GR solutions in the  higher curvature theory framework. Several works have discussed this line of reasoning, for example, \cite{Lu, Mignemi, Bueno}. Many issues devoted to exact solutions in modified gravity theories, that is, first of all, higher curvature theories, were studied, see for a review \cite{Capozziello}. In particular, one of classes of solutions to study is that one where causality is broken down. In fact, the GR is infested by geometries that allow \textit{a priori} to produce a time machine. These pathological space-times present the so-called Closed Time-like Curves (CTCs), as a consequence a traveler moving along such curves can come back to his own past leading to controversial issues, for example, causality paradoxes and time travel. The best-known solutions containing CTCs in GR are Van Stockum \cite{VS}, Gott \cite{Gott} and  G\"{o}del \cite{Godel} time machines.   Further, a generalization of the G\"{o}del solution has been found out, such metrics were denominated by G\"{o}del-type metrics \cite{Reb}. A fundamental feature of these metrics is the possibility of eliminating the CTCs for determined values of their parameters.   	           
         
 Several results have been obtained with respect to the causality aspects of the G\"{o}del-type metrics in frameworks other than GR, such as verification of consistency of these metrics and proof of possibility of existence of completely causal solutions within such theories as the $f(R)$ gravity, Horava-Lifshitz gravity, Brans-Dicke gravity, Chern-Simons modified gravity and $f(R,T)$ gravity \cite{Reb,ourGodel,PP,AF}. Our aim in this paper is treating the G\"{o}del-type metrics in one more higher curvature theory framework, more precisely in $f(R,Q)$ gravity, with $Q=R_{\mu\nu}R^{\mu\nu}$. Many issues related with behavior of the known GR solutions, such as different types of black holes, including Schwarzshild and Reissner-Nordstr\"{o}m ones, and the possible singularity-free black holes, impacts of torsion and wormholes, within this theory were studied in \cite{Olmo}. Several other studies have been also developed: an attempt to find ghost and singularity free theories of gravity has been carried out \cite{Bis}; and in cosmological scenarios, it was found that the Big Bang singularity is replaced by a cosmic bounce in isotropic and anisotropic universes filled with standard sources of matter and radiation \cite{Bar}. To continue studies of this theory, it is necessary to examine causality features, i.e., the likely parameters of the theory allowing CTCs or not as well as their unfolding.  
 
The structure of the paper looks like follows. In the section 2, we review the G\"{o}del-type metrics, resenting their classifications and discussing related causality issues. The section 3 is devoted to deriving of equations of motion in $f(R,Q)$ gravity. In the section 4, we verify the consistency of the G\"{o}del-type metric within the $f(R,Q)$ gravity, and in the section 5 we discuss the causality for the solutions we obtain the matter sources necessary to achieve the causality. Finally, in the section 6 we summarize our results.   

\section{G\"{o}del-type metrics}

In this section we present a brief review on the  known properties of the G\"{o}del-type metrics, more precisely those ones homogeneous in Space-Time (ST-homogeneous), as well as their causality features necessary for further purposes (more details can be found in the papers \cite{Reb}).

Such a metrics are solutions of Einstein field equations and have as their principal property the exhibiting the so-called closed time-like curves. Perhaps, the best-known G\"{o}del-type ST-homogeneous example is the G\"{o}del metric itself. The G\"{o}del universe \cite{Godel} (originally it was proposed as a rotating cosmological model) is generated by non-vanishing cosmological constant $\Lambda$ and  dust-like matter with the density $\rho$ taken as matter sources. The line element in the G\"{o}del spacetime is defined by:
\begin{equation}
ds^2=[dt+H(x)dy]^2-D^{2}(x)dy^{2}-dx^{2}-dz^{2},
\label{Godel}
\end{equation}
where the functions $H(x)$ and $D(x)$ look like
\begin{equation}
H(x)=e^{mx},\quad D(x)=\frac{e^{mx}}{\sqrt{2}},
\label{cond}
\end{equation}
whose parameters of solutions are related with the matter content through the relations:
\begin{equation}
\begin{split}
& m^2=2\omega^2=\kappa\rho;\\
& \Lambda=-\kappa\frac{\rho}{2}
\end{split}
\end{equation}
where $\kappa$ is the Einstein constant and $\omega$ is the vorticity of matter. The G\"{o}del-type metrics are generalizations of the metric (\ref{Godel}, \ref{cond}). In these metrics, the line element in cylindrical coordinates is given by:
 \begin{equation}
 ds^2=[dt+H(r)d\theta]^2-D^{2}(r)d\theta^{2}-dr^{2}-dz^{2},
 \label{metric}
 \end{equation}
where the functions $H(r)$ and $D(r)$ satisfy the following conditions  for the ST-homogeneous case \cite{Reb},
\begin{equation}
\begin{split}
&\frac{H^{'}(r)}{D(r)}=2\omega,\\
&\frac{D^{''}(r)}{D(r)}=m^{2},
\end{split}
\label{ST}
\end{equation}
where the prime denotes the derivative with respect to $r$.  The $m^2$ and $\omega$ are constants characterizing completely the properties of the ST-homogeneous G\"{o}del-type metrics. Their values can be:  $\omega\neq0$ and $-\infty \leq m^2 \leq \infty$. From now we will consider only ST-homogeneous  G\"{o}del-type metrics. Concerning the solutions of Eqs.(\ref{ST}) it is known that there exist three distinct classes of a non-degenerate ($\omega\neq 0$) G\"{o}del-type metrics characterized by the sign of $m^2$, namely  (see also \cite{Reb}):

\begin{itemize}
\item \textit{hyperbolic class}: $m^2>0$, $\omega\neq 0$:
\begin{equation}
\begin{split}
&H(r)=\frac{2\omega}{m^2}[\cosh(mr)-1],\\
&D(r)=\frac{1}{m}\sinh(mr),\\
\end{split}
\end{equation}
\item \textit{trigonometric class}: $-\mu^2=m^2<0$, $\omega\neq 0$:
\begin{equation}
\begin{split}
&H(r)=\frac{2\omega}{\mu^2}[1-\cos(\mu r)],\\
&D(r)=\frac{1}{\mu}\sin(\mu r),\\
\end{split}
\label{trigo}
\end{equation}
\item \textit{linear class}: $m^2=0$, $\omega\neq 0$:
\begin{equation}
\begin{split}
&H(r)=\omega r^2,\\
&D(r)=r.\\
\end{split}
\label{linear}
\end{equation}
\end{itemize}

The G\"{o}del metric is recovered when $m^2=2\omega^2$, thus, it belongs to the hyperbolic class (\ref{Godel}).

 


   
With regard to the causality violation, the circle defined  by $C=\lbrace(t,r,\theta,z); \, t, r, z= \mbox{const.}, \theta \in [0, 2\pi]\rbrace$, is a CTC if $G(r)$ becomes negative for a range of $r$-values ($r_1 <r<r_2$) \cite{Reb}, where $G(r)=D^2 (r)-H^2 (r)$. For the linear class $m^2=0$, one non-causal region exists for $r>r_{c}$, where $r_{c}=1/\omega$ is the critical radius ($r$-value splitting up the causal and non-causal regions). 
  For $m^2=-\mu^2$, the trigonometric class, there is an infinite sequence of alternating causal and non-causal regions. The hyperbolic class ($m^2>0$) can be separated into two depending on $m^2$: the first one occurs for $0< m^2<4\omega^2$, where there is one non-causal region for $r>r_{c}$, with the critical radius $r_c$ given by
\begin{equation}
\sinh^2\bigg(\frac{m r_{c}}{2}\bigg)=\bigg(\frac{4\omega^2}{m^2}-1 \bigg)^{-1}.
\label{rc}
\end{equation}
 The second one occurs when $m^2\geq 4\omega^2$, in this case there is no breakdown of causality and, thus, no occurrence of CTCs (so, this case is completely causal).

\section{The $f(R,Q)$ gravity}

The $f(R,Q)$ gravity action is described by the action (see f.e. \cite{frq,Olmo}):
 \begin{equation}
 S=\frac{1}{2\kappa}\int d^{4}x\,\sqrt{-g}f(R,Q)+S_{mat}(g_{\mu\nu},\psi), 
\label{CSMG} 
 \end{equation}
where $f(R,Q)$ is an arbitrary function of the Ricci scalar $R$ and the curvature invariant defined by $Q=R_{\mu\nu}R^{\mu\nu}$. The matter is minimally coupled to gravity via the matter action $S_{mat}$, $\kappa=8\pi G$ and $g$ is the determinant of metric tensor $g_{\mu\nu}$. In order to get the field equations, in the metric approach, it is necessary to vary the action with respect to $g_{\mu\nu}$. Proceeding in this way we obtain
 \begin{equation}
 \begin{split}
 \delta S=\frac{1}{2\kappa}\int d^{4}x\,\bigg[\delta\sqrt{-g}\,f(R,Q)+\sqrt{-g}\,\delta f(R,Q)\bigg]+\delta S_{mat}(g_{\mu\nu},\psi);
 \end{split}
\label{action}
 \end{equation}
where  
 \begin{eqnarray}
\delta\sqrt{-g}&=&-\frac{\sqrt{-g}}{2}g_{\mu\nu}\delta g^{\mu\nu},\label{iden}\\
\delta f(R,Q)&=&f_{R}\delta R + f_{Q}\delta Q.
 \end{eqnarray}
 with $f_{R}\equiv\dfrac{\partial f}{\partial R}$ and $f_{Q}\equiv\dfrac{\partial f}{\partial Q}$. Using that $\delta R =\delta\big(g^{\mu\nu}R_{\mu\nu}\big)$ and $\delta Q =\delta \big(R_{\mu\nu}R^{\mu\nu}\big)$ we obtain
 \begin{equation}
\begin{split}
\delta R &=R_{\mu\nu}\delta g^{\mu\nu}+g^{\mu\nu}\big(\nabla_{\lambda}\delta\Gamma^{\lambda}_{\mu\nu}-\nabla_{\mu}\delta\Gamma^{\lambda}_{\lambda\nu}\big),
\end{split}
\label{iden1}
\end{equation}
where the Palatini identity has been used  and 
 \begin{equation}
\begin{split}
\delta Q =\delta R_{\mu\nu}R^{\mu\nu}&=2R^{\mu\nu}\delta R_{\mu\nu}+2R_{(\mu}^{\beta}R_{\nu)\beta}\delta g^{\mu\nu}.
\end{split}
\label{iden2}
\end{equation}
 
Putting (\ref{iden},\ref{iden1},\ref{iden2}) into (\ref{action}), we find
\begin{eqnarray}
\delta S&=&\frac{1}{2\kappa}\int d^{4}x\,\bigg[\sqrt{-g}\delta g^{\mu\nu}\bigg(f_{R}R_{\mu\nu}-\frac{f}{2}g_{\mu\nu}+2f_{Q}R_{(\mu}^{\beta}R_{\nu)\beta}+\Lambda g_{\mu\nu}\bigg)\label{eq15}\\
&+&\sqrt{-g}f_{R}g^{\mu\nu}\big(\nabla_{\lambda}\delta\Gamma^{\lambda}_{\mu\nu}-\nabla_{\mu}\delta\Gamma^{\lambda}_{\lambda\nu}\big)+2\sqrt{-g}f_{Q}R^{\mu\nu}\big(\nabla_{\lambda}\delta\Gamma^{\lambda}_{\mu\nu}-\nabla_{\mu}\delta\Gamma^{\lambda}_{\lambda\nu}\big)\bigg]+\delta S_{mat}(g_{\mu\nu},\psi).\nonumber
\end{eqnarray}
Integrating by parts the second and the third term in eq. (\ref{eq15}), and eliminating the boundary terms, the field equations become
\begin{equation}
\begin{split}
& f_{R}R_{\mu\nu}-\frac{f}{2}g_{\mu\nu}+2f_{Q}R_{(\mu}^{\beta}R_{\nu)\beta}+g_{\mu\nu}\square f_{R}-\nabla_{(\mu}\nabla_{\nu)}f_{R}+\\
&+\square\big(f_{Q}R_{\mu\nu}\big)-2\nabla_{\lambda}\big[\nabla_{(\mu}\big(f_{Q}R^{\lambda}_{\nu)}\big)\big]+g_{\mu\nu}\nabla_{\alpha}\nabla_{\sigma}\big(f_{Q}R^{\alpha\sigma}\big)=\kappa T_{\mu\nu}^{(\mbox{m})},
\label{FE1}
\end{split}
\end{equation}
where $T_{\mu\nu}^{(\mbox{m})}=-\dfrac{2}{\sqrt{-g}}\dfrac{\delta(\sqrt{-g} \mathcal L_{\mbox{m}})}{\delta g^{\mu\nu}}$ is the energy-momentum tensor of matter and $\square=\nabla_{\mu}\nabla^{\mu}$ is the covariant d'Alembertian operator. We use the following conventions: the Riemann tensor is $R^{\alpha}_{\,\,\,\mu\beta\nu}=\partial_{\beta}\,\Gamma^{\alpha}_{\nu\mu}-\partial_{\nu}\,\Gamma^{\alpha}_{\beta\mu}-\Gamma^{\alpha}_{\rho\nu}\Gamma^{\rho}_{\beta\mu}+\Gamma^{\alpha}_{\rho\beta}\Gamma^{\rho}_{\nu\mu}$ and for the Ricci tensor is $R_{\mu\nu}=R^{\alpha}_{\,\,\,\mu\alpha\nu}$. Here, we use small Greek letters for coordinate indices running from 0 to 3 and adopt a Lorentzian signature $(+,-,-,-)$.

These field equations can be written in Einstein-like form, i.e.,
\begin{equation}
R_{\mu\nu}-\frac{1}{2}Rg_{\mu\nu}=\kappa_{eff} T^{(\mbox{m})}_{\mu\nu}+T_{\mu\nu}^{eff},
\label{eff}
\end{equation}
where $\kappa_{eff}=\dfrac{\kappa}{f_R}$ and
\begin{equation}
\begin{split}
T_{\mu\nu}^{eff}&=\frac{1}{f_{R}}\bigg(
-\frac{1}{2}Rg_{\mu\nu}f_{R}+\frac{f}{2}g_{\mu\nu}-2f_{Q}R_{(\mu}^{\beta}R_{\nu)\beta}-g_{\mu\nu}\square f_{R}+\nabla_{(\mu}\nabla_{\nu)}f_{R}-\\
&-\square\big(f_{Q}R_{\mu\nu}\big)+2\nabla_{\lambda}\big[\nabla_{(\mu}\big(f_{Q}R^{\lambda}_{\nu)}\big)\big]-g_{\mu\nu}\nabla_{\alpha}\nabla_{\sigma}\big(f_{Q}R^{\alpha\sigma}\big)
\bigg),
\label{effdef}
\end{split}
\end{equation}
is the effective energy-momentum tensor.

It is more convenient to write the field equations in the trace-reversed form, thus taking the trace of (\ref{eff})
\begin{equation}
\begin{split}
R&=-\bigg(\kappa_{eff} T^{(\mbox{m})}+T^{eff}
\bigg),
\label{trace}
\end{split}
\end{equation}
where $T^{(\mbox{m})}=g^{\mu\nu}\,T^{(\mbox{m})}_{\mu\nu}$ and  $T^{eff}=g^{\mu\nu}\,T^{eff}_{\mu\nu}$.

Using this result, one can write (\ref{eff}) as
  \begin{equation}
R_{\mu\nu}=\kappa_{eff}\bigg(T^{(\mbox{m})}_{\mu\nu}-\frac{1}{2}g_{\mu\nu}T^{(\mbox{m})}\bigg)+\bigg(T_{\mu\nu}^{eff}-\frac{1}{2}g_{\mu\nu}T^{eff}\bigg).
\end{equation}
 We note that within our studies, the energy-momentum tensor of the matter is conserved. Indeed, it is possible to verify that the divergence of the r.h.s. of the Eq. (\ref{FE1}) vanishes, and the matter we consider throughout this paper is usual (relativistic fluid, scalar or electromagnetic field). From the physical viewpoint, it is related with the fact that within our studies, a space-time is suggested to be homogeneous.

In the next section we deal with the problem of the causality violation in the $f(R,Q)$ theory using the G\"{o}del-type metrics.

\section{G\"{o}del-type metrics in $f(R,Q)$ gravity}

To study the equations of motion in our theory, for the sake of simplicity, we will use the Cartan formalism.  Following its principles, we define a Lorentzian manifold $M$, a local section of its orthonormal frame bundle $F(M)$ with structure group $SO(3,1)$ (the frame bundle is defined by $F(M)=\bigcup\limits_{p \in M} F_{p}$, where $F_p$ is the set of all orthonormal frames ${e_{A}}$ defined at each point $p$ in $M$, thus it is a fiber of $F(M)$ in $p$) is a orthonormal frame field, also called a tetrad or vierbein, $e_{A}(x)=e^{\,\,\,\mu}_{A}(x)\partial_{\mu}$ whose set of such a vectors forms a basis for the tangent space $T_{p}(M)$ at each point $p$ in $M$. Equivalently, we can define the dual frame field or co-frame field $\theta^{A}(x)=e^{A}_{\,\,\,\mu}(x)dx^\mu$ where the set of these vectors is a basis for the cotangent space $T_{p}^{*}(M)$. The duality condition $e_{A}(\theta^{B})=\delta^{B}_{A}$ leads to $e^{\mu}_{\,\,\,A}e^{A}_{\,\,\,\nu}=\delta^{\mu}_{\nu}$ and $e^{\mu}_{\,\,\,A}e^{B}_{\,\,\,\mu}=\delta^{B}_{A}$. Here, capital Latin letters label Lorentz indices and run from 0 to 3.
For the G\"{o}del-type manifolds given by (\ref{metric}) we can define a local Lorentz (orthonormal) co-frame such that
\begin{eqnarray}
\theta^{(0)}&=& dt+ H(r)d\theta;\nonumber\\
\theta^{(1)}&=& dr;\nonumber\\
\theta^{(2)}&=&D(r) d\theta;\nonumber\\
\theta^{(3)}&=& dz,
\label{TB}
\end{eqnarray}
where $ds^2=\eta_{AB}\theta^{A}\theta^{B}$, with $\eta_{AB}=diag(+1,-1,-1,-1)$ being the Minkowski metric. In this co-frame, the field equations become
\begin{equation}
R_{AB}=\kappa_{eff}\bigg(T^{(\mbox{m})}_{AB}-\frac{1}{2}\eta_{AB}T^{(\mbox{m})}\bigg)+\bigg(T_{AB}^{eff}-\frac{1}{2}\eta_{AB}T^{eff}\bigg).
\label{32}
\end{equation}
In the Lorentz co-frame (\ref{TB}), the non-vanishing components of Ricci tensor are $R_{(0)(0)}=2\omega^2$, $R_{(1)(1)}=R_{(2)(2)}=2\omega^2-m^2$. Note that, all the components are constants. The Ricci scalar is $R=2(m^2-\omega^2)$, and $Q=2m^2(m^2-4\omega^2)+12\omega^4$, both are also constants.

Since the $R$ and $Q$ scalars are constants for the G\"{o}del-type metrics, the eq.(\ref{effdef}) may be simplified. Then
\begin{eqnarray}
T_{\mu\nu}^{eff}&=&\frac{1}{f_{R}}\bigg(
-\frac{1}{2}Rg_{\mu\nu}f_{R}+\frac{f}{2}g_{\mu\nu}-2f_{Q}R_{(\mu}^{\beta}R_{\nu)\beta}-f_{Q}\square R_{\mu\nu}+2f_{Q}\nabla_{\lambda}\nabla_{(\mu}R^{\lambda}_{\nu)}-f_{Q}g_{\mu\nu}\nabla_{\alpha}\nabla_{\sigma}R^{\alpha\sigma}
\bigg)\nonumber\\
&=&\frac{1}{f_{R}}\bigg(-\frac{1}{2}Rg_{\mu\nu}f_{R}+\frac{f}{2}g_{\mu\nu}+2f_{Q}R_{\mu\lambda\theta\nu}R^{\lambda\theta}-f_{Q}\square R_{\mu\nu}\bigg),
\label{24}
\end{eqnarray} 
where have been used the fact that derivatives of $R$ and $Q$ are null and the following identities 
\begin{equation}
\begin{split}
\nabla_{\rho}\nabla_{\nu}R_{\mu}^{\rho}&=\frac{1}{2}\nabla_{\nu}\nabla_{\mu}R+R_{\mu\lambda\theta\nu}R^{\lambda\theta}+R^{\lambda}_{\mu}R_{\nu\lambda},\\
\nabla_{\mu}\nabla_{\nu}R^{\mu\nu}&=\frac{1}{2}\square R.
\end{split}
\end{equation}
Note that the effective energy-momentum  tensor in the co-frame (\ref{TB}) is given by $T_{A B}^{eff}=e^{\mu}_{A}e^{\nu}_{B}T_{\mu\nu}^{eff}$. Furthermore, if only the higher-order derivative term,  i.e., $X_{\mu\nu}=-f_{Q}\square R_{\mu\nu}$ is considered the non-vanishing components in the local Lorentz co-frame (\ref{TB}) for the G\"{o}del-type metrics are
 \begin{equation}
\begin{split}
X_{(0)(0)}&=4f_{Q}\omega^2(4\omega^2-m^2),\\
X_{(1)(1)}=X_{(2)(2)}&=2f_{Q}\omega^2(4\omega^2-m^2).
\end{split}
\end{equation}
Thus, differently from \cite{Clifton}, we found that the only possibility of the field equations to reduce to second order is $m^2=4\omega^2$ for all $f_{Q}\neq 0$, it leads to the vanishing of the higher-order derivative term and, consequently, avoiding possible instabilities. It is evident that if a tensor is null in a particular frame it is null for any other frames.

 Since the trace of $T_{\mu\nu}^{eff}$ reduces to $T^{eff}=\dfrac{1}{f_R}\bigg(-2Rf_R +2f-2f_{Q}R^{\alpha\beta}R_{\alpha\beta}\bigg)$ that, in turn, by substituting in (\ref{trace}) it lead us to a constraint equation, namely,
\begin{equation}
f_{R}R+2f_{Q}Q-2f=\kappa T^{(\mbox{m})},
\label{trace1}
\end{equation}
such equation is indeed an algebraic equation which relates the matter content to geometric quantities.

An important ingredient we must implement is the matter content, in order to obtain new results we, besides of a perfect fluid, will use a massless scalar field. The perfect fluid has density $\rho$ and pressure $p$, its energy-momentum tensor is given by $T^{(pf)}_{AB}=(p+\rho)u_{A}u_{B}-p\eta_{AB}$, in the local Lorentz co-frame (\ref{TB}), thus
\begin{equation}
T^{(pf)}_{\ (0)(0)}=\rho, \quad\quad T^{(pf)}_{\ (1)(1)}=T^{(pf)}_{\ (2)(2)}=T^{(pf)}_{\ (3)(3)}=p,
\end{equation} 
where we have defined the 4-velocity of a point of fluid $u^A=e^{A}_{0}=\delta^{A}_{0}$. Now, let us treat the massless scalar field $\psi$ which in its turn satisfies the Klein-Gordon equation $\square\psi=\eta^{AB}\big(\nabla_{A}\nabla_{B}\psi+\omega^{C}_{\ BA}\nabla_{C}\psi\big)=0$. Due to the symmetry of the metric we take the gradient of $\psi$ in $z$-direction, in other words, $\nabla_A \psi=e^{\ \mu}_{A}\nabla_{\mu}\psi=b\hat{z}$ implying $\psi=b(z-z_{0})$, where $b$ and $z_{0}$ are constants. Such a choice leads to the non-vanishing components of the energy-momentum tensor $T^{sf}_{AB}=\nabla_{A}\psi\nabla_{B}\psi-\frac{1}{2}\eta_{AB}\nabla_{C}\psi\nabla^{C}\psi$ for the scalar field in the co-frame (\ref{TB}) are
\begin{equation}
T^{(sf)}_{\ (0)(0)}=T^{(sf)}_{\ (3)(3)}=\frac{1}{2}b^2, \quad\quad T^{(sf)}_{\ (1)(1)}=T^{(sf)}_{\ (2)(2)}=-\frac{1}{2}b^2,
\end{equation}
as a consequence, $T^{\mbox{(m)}}_{\ AB}=T^{(pf)}_{\ AB}+T^{(sf)}_{\ AB}$ and, thus, we have
\begin{equation}
T^{\mbox{(m)}}_{\ (0)(0)}=\rho+\frac{1}{2}b^2, \quad T^{\mbox{(m)}}_{\ (1)(1)}=T^{\mbox{(m)}}_{\ (2)(2)}=p-\frac{1}{2}b^2,  \quad T^{\mbox{(m)}}_{\ (3)(3)}=p+\frac{1}{2}b^2.
\label{35} 
\end{equation}

Similarly, the non-vanishing components energy-momentum tensor $T^{eff}_{\ AB}=e^{\ \mu}_{A}e^{\ \nu}_{B}T^{eff}_{\ \mu\nu}$, in (\ref{TB}), are 
\begin{equation}
\begin{split}
T^{eff}_{\ (0)(0)}&=\frac{1}{2}\frac{2(\omega^2-m^2)f_R+16\omega^2(3\omega^2-m^2)f_Q+f}{f_{R}};\\
T^{eff}_{\ (1)(1)}&=T^{eff}_{\ (2)(2)}=-\frac{1}{2}\frac{2(\omega^2-m^2)f_R+4(6\omega^2 m^2-m^4-12\omega^4)f_Q+f}{f_{R}};\\
T^{eff}_{\ (3)(3)}&=-\frac{1}{2}\frac{2(\omega^2-m^2)f_R+f}{f_{R}}.
\label{MS}
\end{split}
\end{equation}

As discussed in section III, the energy-momentum tensor of the matter is conserved. As a consequence the field equations, more precisely Eq. (\ref{eff}), lead to the constraint $\nabla_{\mu}T^{\mu\nu}_{(eff)}=0$, in other words, the effective energy-momentum tensor is conserved as well. Hence, we must check whether the G\"{o}del-type metrics satisfy such a constraint, so by means of the straightforward calculation we get:
\begin{equation}
\begin{split}
\nabla_{\mu}T^{\mu\nu}_{(eff)}&=\nabla_{\mu}\big(e^{\mu}_{A}e^{\nu}_{C}T^{A C}_{(eff)}\big)\\
&=e^{\nu}_{C}\big[\nabla_{A}T^{A C}_{(eff)}+\big(\nabla_{\mu}e^{\mu}_{A}\big)T^{A C}_{(eff)}\big]+e^{\mu}_{A}\big(\nabla_{\mu}e^{\nu}_{C}\big)T^{A C}_{(eff)}=0,
\end{split}
\end{equation}
multiplying by $e^{B}_{\nu}$ one finds the expression in a non-holonomic frame, namely,
\begin{equation}
\begin{split}
\nabla_{A}T^{A B}_{(eff)}+\omega^{C}_{\ A C}T^{A B}_{(eff)}+\omega^{B}_{\ C A}T^{A C}_{(eff)}=0,
\end{split}
\label{NHframe}
\end{equation}
where $\omega^{B}_{\ C A}$ are the components of spin connection and we have used the same definitions of \cite{PP}. Finally, by direct replacement Eq.(\ref{MS}) into Eq.(\ref{NHframe}) one finds that the requirement is fulfilled.

Thus, the field equations in Lorentz co-frame (\ref{TB}) for the G\"{o}del-type metrics with matter content (\ref{MS}) are given by
\begin{eqnarray}
4\omega^2 f_R-2\kappa\rho-f-16f_Q\omega^2(3\omega^2-m^2)-\kappa b^2&=&0,\\
2f_R(2\omega^2-m^2)-2\kappa p-4f_Q(12\omega^4+m^4-6\omega^2 m^2)+f+\kappa b^2&=&0,\label{third1}\\
f-2\kappa p-\kappa b^2&=&0,\label{third}
\end{eqnarray}
or, as is the same,
\begin{eqnarray}
\kappa b^2&=&(m^2-2\omega^2)f_R+2(m^4-6\omega^2 m^2+12\omega^4)f_Q,\label{55}\\
\kappa p&=&\frac{1}{2}f-\frac{1}{2}(m^2-2\omega^2)f_R-(m^4-6\omega^2 m^2+12\omega^4)f_Q,\\
\kappa \rho&=&-\frac{1}{2}f-\frac{1}{2}(m^2-6\omega^2)f_R-(36\omega^4-14\omega^2 m^2+m^4)f_Q\label{57}.
\end{eqnarray}

Now, let us treat the general features of the field equations for this, it is worth pointing out some special situations: the first one takes place for G\"{o}del solution ($m^2=2\omega^2$), that presents CTCs. In this case $b$ depends on $f_Q$, as may be seen from (\ref{55}). Explicitly, this statement can be verified by substituting $m^2=2\omega^2$ into eqs.(\ref{55}-\ref{57}) leading to
\begin{eqnarray}
\kappa b^2&=&8\omega^4f_Q,\label{55a}\\
\kappa p&=&\frac{1}{2}f-4\omega^4f_Q,\\
\kappa \rho&=&-\frac{1}{2}f+2\omega^2f_R-12\omega^4f_Q\label{57a},
\end{eqnarray}
whose set up is univocally determined for some specified $f(R,Q)$. In order to determine the causality features of the G\"{o}del solution given by eqs.(\ref{55}-\ref{57}), it is necessary to consider the eq.(\ref{rc}), that defines the critical radius $r_c$, taking $m^2=2\omega^2$, i.e, the $r_c$ is given by 
\begin{equation}
r_c=\frac{2}{m}\sinh^{-1} (1)=2\sinh^{-1}(1) \sqrt{\dfrac{f_R}{\kappa(\rho+p+2b^2)}},
\end{equation}
where we have used eqs. (\ref{55}-\ref{57}) in the last step. Notice that $r_c$ depends only on matter content and $f_R$. We found that $b\neq 0$ implies $f_Q\neq 0$, and our result generalized that one from \cite{JSantos}. In particular, when $f(R,Q)=f(R)$, we obtain the results found in \cite{JSantos} where  only G\"{o}del solution is possible for pure perfect fluid, i.e., $b=0$.

Other relevant G\"{o}del-type solutions are the linear and trigonometric classes which both are compatible with the existence of CTCs. In the special case of the linear class ($m^2=0$), the equations of motion (\ref{55}-\ref{57}) become 
\begin{eqnarray}
\kappa b^2&=&24\omega^4f_Q-2\omega^2f_R,\label{58}\\
\kappa p&=&\frac{1}{2}f+\omega^2f_R-12\omega^4f_Q,\\
\kappa \rho&=&-\frac{1}{2}f+3\omega^2f_R-36\omega^4f_Q\label{60}.
\end{eqnarray}
From the eqs(\ref{58}-\ref{60}) we found a relation between the matter sources: $p+\rho=-2b^2$. Furthermore, the $r_c$ for linear class is given by 
\begin{equation}
r_c=\left[\frac{f_R}{24 f_Q}\left(1+\sqrt{1+24\kappa f_Q\bigg(\frac{b}{f_R}\bigg)^{2}}\,\right)\right]^{-1/2},
\end{equation}
for all $f_R> 0$ and $f_Q> 0$.

In the subsection below, we examine particular matter sources in the $f(R,Q)$ framework. 
	
\subsection{Vacuum solutions}

In contrast to GR \cite{Reb}, the $f(R,Q)$ gravity admits G\"{o}del-type vacuum solutions. In such case, it is necessary to add a cosmological constant $\Lambda$ into field equations, that can be made through redefinition $f(R,Q)\rightarrow f(R,Q)-2\Lambda$. Having this in mind, the eqs. (\ref{trace}) and (\ref{third}) reduce to
\begin{equation}
\begin{split}
0=&f_{R}R+2f_{Q}Q\\
=&f_R (m^2-\omega^2)+f_Q (12\omega^4+2m^4-8\omega^2 m^2)
\label{eq1}
\end{split}
\end{equation} 
where $R=2(m^2-\omega^2)$ and $Q=2m^2(m^2-4\omega^2)+12\omega^4$ have been used. On the other hand, the eq.(\ref{third1}) becomes
\begin{equation}
f_{R}(2\omega^2-m^2)-f_{Q}(24\omega^4+2m^4-12\omega^2 m^2)=0.
\label{eq2}
\end{equation}
Therefore, combining the eqs. (\ref{eq1}) and (\ref{eq2}) remain
\begin{equation}
f_R +4f_Q m^2 -12f_Q \omega^2=0.
\label{frfq}
\end{equation}
Recalling that by taking $f_Q=0$ implies, necessarily, $f_R=0$ as a consequence the above equation turns out not to be consistent in the GR framework ($f_R=1$ and $f_Q=0$), therefore, there is not G\"{o}del-type vacuum solution in GR in according with \cite{Reb}. 
An important particular case corresponds to the conformally flat space and CTC-free $(m^2=4\omega^2)$ which, in turn, it leaves the field equations of second order-derivative, for this situation the eq.(\ref{frfq}) reduces to $f_R+4\omega^2 f_Q=0$ (we note that our results are in disagreement with \cite{Clifton}, but we believe that there is a some error there).

The eq.(\ref{frfq}) must be solved specifying both $f_R$ and $f_Q$ which are evaluated at $R=2(m^2-\omega^2)$ and $Q=2m^2(m^2-4\omega^2)+12\omega^4$, thus generating an algebraic equation of the form $m=m(\omega)$. To do so, let us pick up a specific theory, for instance, $f=R+\alpha R^2+\beta Q$, using it into the former equation we found
\begin{equation}
4 m^2=\frac{4\omega^2 (\alpha+3\beta)-1}{\alpha+\beta}.
\label{vc}
\end{equation}
Note that the theory aforementioned, in particular, admits the three class of G\"{o}del-type metrics. When $m^2=0$ (linear class) the eq.(\ref{vc}) reduces to
\begin{equation}
\omega^2=\frac{1}{4(\alpha+3\beta)},
\end{equation}
 with the parameters satisfying the following conditions: $\alpha+\beta\neq 0$ and $\alpha+3\beta\neq 0$. The trigonometric class ($m^2<0$) is recovered when 
\begin{equation}
0<\omega^2<\frac{1}{4(\alpha+3\beta)},
\end{equation}
with $\alpha+\beta>0$. Indeed, there is another possibility, however unphysical because $\alpha\rightarrow 0$ and $\beta\rightarrow 0$ lead to $\omega^2\rightarrow\infty$.
  For the hyperbolic class ($m^2>0$) we have
 \begin{equation}
0<\omega^2<\frac{1}{4(\alpha+3\beta)},
\label{vc1}
\end{equation}
with $\alpha+\beta<0$.
An interesting case of the hyperbolic class corresponds to the geometry $m^2=\omega^2$ such a situation implies $R=0$, thus we may determine a range of validity for the parameters. On the other hand, by means of eq. (\ref{vc}) we have $\omega^2=\dfrac{1}{8\beta}>0$ that, in turn, must be within the range (\ref{vc1}). However, by replacing in the eq. (\ref{vc1}) we found $\alpha+\beta<0$ which is clearly in accordance, thus the space $m^2=\omega^2$ is a vacuum solution of $f=R+\alpha R^2+\beta Q$.
  
Particularly, we found a completely causal G\"{o}del-type vacuum solutions that there is not analogue in the GR framework. The solutions are obtained by imposing the condition $m^2\geq 4\omega^2$ in eq.(\ref{vc}) leading to
\begin{equation}
\begin{split}
&\omega^2\geq \frac{1}{4|3\alpha+\beta|},\\
&f=2\Lambda,
\label{UK}
\end{split}
\end{equation}
where $3\alpha+\beta<0$ and it is required $\alpha-\beta>0$ so that the eqs.(\ref{vc1}-\ref{UK}) are in agreement.  

The first completely causal solution takes place for $m^2=4\omega^2$, where $r_{c}\rightarrow\infty$, that corresponds to equality in eq.(\ref{UK}), thus the solution is:
\begin{eqnarray}
\Lambda&=&\frac{3}{2}\omega^2,\\
m^2&=&4\omega^2=\frac{1}{|3\alpha+\beta|},
\label{mm}
\end{eqnarray}
where $3\alpha+\beta<0$ due to the positivity of $\omega^2$, and the cosmological constant is positively definite, this case have been treated in \cite{Accioly}, however our result differs by an additional negative sign in both equations.

\section{Causal Solutions in the presence of matter sources}

In this section we treat the possibility of the existence of causal solutions for the matter content composed by perfect fluid and scalar field already aforementioned above. In order to evaluate causal solutions it is necessary that the condition $m^2\geq 4\omega^2$ be satisfied. Taking this into account, it is possible to determine constraints on the $f(R,Q)$ theory. 

The causality features become clearer by writing the field equations (\ref{55}-\ref{57}) into the form
\begin{eqnarray}
2f_R&=&\frac{\kappa(m^4-6\omega^2m^2+12\omega^4)(p+\rho)}{\omega^2 m^2(4\omega^2-m^2)}+\frac{\kappa(4\omega^2-m^2)(6\omega^2-m^2)b^2}{\omega^2 m^2(4\omega^2-m^2)},\label{69}\\
4f_Q&=&\frac{\kappa(2\omega^2-m^2)(p+\rho)}{\omega^2 m^2(4\omega^2-m^2)}+\frac{\kappa b^2}{\omega^2 m^2},\label{70}
\end{eqnarray}
for all $m^2\neq 4\omega^2$ and $m^2\neq 0$.
  In particular, the first causal solution arises when $m^2=4\omega^2$ it is evident that the above equations do not apply, thus we should use eqs.(\ref{55}-\ref{57}) which, in turn, reduce to 
\begin{eqnarray}
\rho+p&=&0,\label{70a}\\
\kappa b^2&=&2\omega^2 f_R+8\omega^4 f_Q\label{71},\\
\kappa(2p+b^2)&=&f\label{72}.
\end{eqnarray}

For the pure perfect fluid case we have the relation $f_R=-4\omega^2 f_Q$ which agree with the results obtained in \cite{Clifton} except for a negative sign (we believe that in \cite{Clifton}, the sign in equations of motion was lost). 
 Particularly, the case $f=R+\alpha R^2+\beta Q$   have been treated in \cite{Accioly}, in this situation the eqs. (\ref{70}-\ref{72}) reduce to
\begin{equation}
\begin{split}
\kappa b^2&=2\omega^2+8\omega^{4}(3\alpha+\beta),\\
m^2=4\omega^2&=\frac{4}{|3\alpha+\beta|},
\end{split}
\end{equation}
according to the results obtained in \cite{Accioly}.
Returning to general case, the scalar field plays a underlying role because there is an arbitrariness in choosing $f_R$ and $f_Q$ wider than in the pure perfect fluid case. It can be verified that eq.(\ref{71}) leads to inequality $f_R>-4\omega^2 f_Q$, in other words, the presence of the scalar field allows a greater arbitrariness on the choice of $f(R,Q)$ function.
  
The other causal solutions are got by imposing the condition $m^2>4\omega^2$ to eqs.(\ref{69}-\ref{70}), so that some requirements must be fulfilled, for this purpose we might split up into three cases:

\begin{itemize}

\item$p+\rho=0$ and $b^2>0$.

This situation implies the following conditions: $$\left\{\begin{array}{rc}
f_R>0, & \mbox{if}\quad 4\omega^2<m^2<6\omega^2,\\ 
f_R=0, & \mbox{if}\quad m^2=6\omega^2,\\
f_R<0, & \mbox{if}\quad m^2>6\omega^2,\\
f_Q>0,& \mbox{everywhere}.\\
\end{array}\right.
$$

\item $p+\rho>0$ and $b^2>0$. 

In this case we have the following conditions:  $$\left\{\begin{array}{rc}
f_Q>0, & \mbox{everywhere},\\ 
f_R<0,&\mbox{if}\quad m^2\geq 6\omega^2.\\
\end{array}\right.
$$

The range corresponding to $4\omega^2<m^2<6\omega^2$ leads to both possibilities $f_R>0$ and $f_R<0$ depending on the relationship between the matter sources. Note that an  interesting particular case of completely causal solution arises for the pure scalar field and $m^2=6\omega^2$ so that $f(R,Q)$ reduces to $f(Q)$.

\item $p+\rho<0$ and $b^2>0$. 

This case is quite different from the former ones. Now, $f_Q$ admits both signs as well as $f_R$ depending on the matter content as can be seen from eqs. (\ref{69}-\ref{70}), apart from within range $4\omega^2<m^2\leq 6\omega^2$ where $f_R>0$.   

\end{itemize}

 In order to obtain an analysis more detailed we take again $f=R+\alpha R^2+\beta Q$ for the case corresponding to $p+\rho=0$ and $b^2>0$. Evidently, we have three possibilities to find solutions  without CTCs: the first one occurs when $4\omega^2<m^2<6\omega^2$, culminating in the following: $f_R>0$, $f_Q>0$ and $f=\kappa b^2$, as above mentioned. The first condition provides us a relation between $\omega$ and $\alpha$, i.e., 
\begin{equation}
\omega^2<\frac{1}{20|\alpha|},
\end{equation}
where $\alpha<0$ while that the second condition implies $\beta>0$, or we can still have 
\begin{equation}
\omega^2>-\frac{1}{20\alpha},
\end{equation}
when $\alpha>0$, i.e., the range corresponding to $4\omega^2<m^2<6\omega^2$ is valid for all $\alpha$ although it is only valid for $\beta>0$. In analogy, the second possibility occurs at the range, $m^2=6\omega^2$, leading to $f_R=0$ and $f_Q>0$, thus 
\begin{equation}
\omega^2=\frac{1}{20|\alpha|},
\end{equation}
where $\alpha<0$ and $\beta>0$ must be satisfied, note that differently to the previous case now $\alpha>0$ is no longer holds. Finally, the last possibility takes place when $m^2>6\omega^2$ whenever $f_R<0$ and $f_Q>0$ hold, and similarly to the previous cases we find a relation for $\omega$ and $\alpha$ given by 
\begin{equation}
\omega^2<\frac{1}{20|\alpha|},
\end{equation}
where $\alpha<0$ and $\beta>0$, on the other hand $\alpha>0$ implies necessarily $f_R>0$, thus it is not valid.

\section{Summary}

The G\"{o}del-type metric within the context of the $f(R,Q)$ gravity has been considered for physically well-motivated matter sources presented by perfect fluid and scalar field. We note that in general, extension of the gravity Lagrangian enriches the structure of possible solutions. This is just the situation occurring in our theory. We verified that the field equations of the $f(R,Q)$ theory reduce to the second-order derivative equations of motion, thus it is ghost-free and CTC-free for the maximum isometry group of the G\"{o}del-type metric ($m^2=4\omega^2$). This means that in this case the $f(R,Q)$ theory is completely stable as well as causal. Furthermore, the necessary conditions for arising all three G\"{o}del-type classes have been found. Indeed, our main result is that, within this theory, there are essentially new solutions, that is, completely causal G\"{o}del-type solutions which are absent in GR. A remarkable result have been the existence of causal vacuum G\"{o}del-type solutions in the presence of non-null cosmological constant, such a solutions have not analogue in GR.
 
Taking into account the matter sources we also found the conditions for existence of completely causal solutions. In particular, when the scalar field is null, we note that our analysis covers both the case of the usual matter, that is, $\rho+p>0$, and the case of the exotic matter, that is, $\rho+p<0$. Therefore, we see that the exotic matter for this case is not required for the existence of completely causal solutions. On the other hand, the inclusion of the scalar field  is of fundamental importance because it permits a wide arbitrariness for the choice of $f_R$ and $f_Q$.   Since the results depend explicitly on the function $f(R,Q)$, as an example, we considered the particular model where $f(R,Q)=R+\alpha R^2+ \beta Q$. By studying this model we classified the possible values of the parameters $\alpha$ and $\beta$ with respect to the possibility of arising CTCs.

To close the paper, we note that the G\"{o}del-type metric describes the rotating Universe, but without taking into account its expansion. Nevertheless, in general, metrics involving rotation play a central role in gravitational physics for many reasons. It is interesting to note some of them, first, the possibility of rotation of the Universe is treated as a rather interesting idea within the cosmological context \cite{BarrowUn}, second, the rotation of the Universe would imply in the presence of the privileged space-time direction, that is, the rotation axis, which clearly signalizes the possibility of the Lorentz symmetry breaking, which makes studies of Lorentz-breaking theories, and, especially, the Lorentz-breaking gravity, to be extremely important. Some results in this direction are presented in \cite{Tasson}.

  
\textbf{Acknowledgments.} This work was partially supported by Conselho Nacional de Desenvolvimento Científico e Tecnológico (CNPq). The work by A. Yu. P. has been partially supported by the CNPq project No. 303783/2015-0.


\begin{thebibliography}{50}

\bibitem{LIGO1} B. P. Abbott et al., 
Phys. Rev. Lett. 116, 061102 (2016), arXiv: 1602.03837.

\bibitem{LIGO2} B. P. Abbott and et al.,
Phys. Rev. Lett. 116, 241103 (2016), arXiv: 1606.04855. 

\bibitem{LIGO3} B. P. Abbott et. al., 
Phys. Rev. X6, 041015 (2016), arXiv:  1606.04856.

\bibitem{Riess} A. G. Riess, et al.  
Astron. J. 116, 1009 (1998), astro-ph/9805201. 

\bibitem{Riess2} A. G. Riess, et al. 
Astron. J. 607, 665 (2004), astro-ph/0402512. 

\bibitem{Perlmutter} S. Perlmutter, et al. 
Astron. J. 517, 565 (1999), astro-ph/9812133.

\bibitem{boomerang} P. de Bernardis, et al. 
Nature 404, 955  (2000), astro-ph/0004404.

\bibitem{boomerang2} A. E. Lange, et al. 
Phys.Rev. D63, 042001 (2001), astro-ph/0005004.
 
\bibitem{WMAP} G. Hinshaw et al. 
Astrophys. J. Suppl. 208, 19 (2013), arXiv:1212.5226 [astro-ph.CO].

\bibitem{LAnderson}L. Anderson et al., 
MNRAS 427, 3435 (2012), arXiv:1203.6594 [astro-ph.CO].


\bibitem{Samushia} L. Samushia, B. A. Reid, M. White, W. J. Percival, A. J. Cuesta, et al., 
Mon. Not. Roy. Astron. Soc. 429, 1514 (2013), arXiv:1206.5309 [astro-ph.CO].

\bibitem{Veltman} G.~'t Hooft and M.~J.~G.~Veltman,
  Ann.\ Inst.\ H.\ Poincare Phys.\ Theor.\ A20, 69 (1974).
  
\bibitem{Stelle} K. S. Stelle, 
Phys. Rev. D 16,  953 (1977).  

\bibitem{Polchinski} J. Polchinski, \textit{String theory. Vol. 2: Superstring theory
and beyond}, Cambridge University Press, Cambridge,
UK, 1998; S.~L.~Li, X.~H.~Feng, H.~Wei and H.~L\"u,
  Eur.\ Phys.\ J.\ C {\bf 77}, 289 (2017),
  arXiv:1612.02069 [hep-th].
	
\bibitem{Konoplya:2016pmh}
  R.~Konoplya and A.~Zhidenko,
  Phys.\ Lett.\ B {\bf 756}, 350 (2016),
  arXiv:1602.04738 [gr-qc].

\bibitem{Lu} H. Lu, A. Perkins, C.N. Pope and K. S. Stelle, 
Phys. Rev. Lett. 114, 171601 (2015), arXiv: 1502.01028.
 
\bibitem{Mignemi} S. Mignemi and D.L. Wiltshire, 
Phys. Rev. D 46, 1475 (1992), hep-ph/9202031.

\bibitem{Bueno} P. Bueno and P. A. Cano, \textit{On black holes in higher-derivative gravities}, arXiv:1703.04625.  

\bibitem{Capozziello} S. Capozziello, M. Francaviglia, Gen. Rel. Grav. 40, 357 (2008), arXiv: 0706.1146;
S. Capozziello, M. de Laurentis, Phys. Rept. 509, 167 (2011), arXiv: 1108.6266. 

\bibitem{VS} W. J. Van Stockum, Proc. R. Soc. Edinburgh A57, 135 (1937).

\bibitem{Gott} J. R. Gott, Phys. Rev. Lett. 66, 1126 (1991). 

\bibitem{Godel} K. Godel, Rev. Mod. Phys. 21, 447 (1949).

\bibitem{Reb} M. J. Reboucas, J. Tiomno, Phys. Rev.  D28, 1251 (1983), Nuovo Cim. B90, 204 (1985); A. F. F. Teixeira, M. Reboucas, J. Aman, Phys. Rev. D32, 3309 (1985), J. Math. Phys. 27, 1370 (1986); M. O. Calvao, M. Reboucas, A. F. F. Teixeira, W. M. Silva, J. Math. Phys. 29, 1127 (1989); M. O. Calvao, I. D. Soares, J. Tiomno, Gen. Rel. Grav. 22, 683 (1990); H. L. Carrion, M. Reboucas, A. F. F. Teixeira, J. Math. Phys. 40, 4011 (1999), gr-qc/9904074; M. J. Reboucas, J. Santos, Phys. Rev. D80, 063009 (2009), arXiv: 0906.5354; J. Santos, M. Reboucas, T. B. R. F. Oliveira, Phys. Rev. 81, 123017 (2010), arXiv: 1004.2501; J. B. Fonseca-Neto, A. Yu. Petrov, M. Reboucas, Phys, Lett. B725, 412 (2013), arXiv: 1304.4675.

\bibitem{ourGodel} C. Furtado, T. Mariz, J. R. Nascimento, A. Yu. Petrov, A. F. Santos, Phys. Rev. D79, 124039 (2009), arXiv: 0906.0554; C. Furtado, J. R. Nascimento, A. Yu. Petrov, A. F. Santos, Phys. Lett. B693, 494 (2010), arXiv: 1004.5106.

\bibitem{PP} J. A. Agudelo, P. J. Porf\'{i}rio, J. B. Fonseca-Neto, J. R. Nascimento,
A. Yu. Petrov and A. F. Santos, Phys. Lett. B762, 96 (2016), arXiv: 1603.07582;
P. J. Porf\'{i}rio, J. B. Fonseca-Neto, J. R. Nascimento,
A. Yu. Petrov, J. Ricardo and A. F. Santos, Phys.Rev. D94, 044044 (2016), arXiv: 1606.00743;   P. J. Porf\'{i}rio, J. B. Fonseca-Neto, J. R. Nascimento,
A. Yu. Petrov,  Phys.Rev. D94, 104057 (2016), arXiv: 1610.01539.

\bibitem{AF} A. F. Santos,  Mod. Phys. Lett. A {\bf 28}, 1350141 (2013); A. F. Santos and C. J. Ferst, Mod. Phys. Lett. A {\bf 30}, 1550214 (2015).   

\bibitem{Olmo} G.~J.~Olmo and D.~Rubiera-Garcia,
  Phys.\ Rev.\ D {\bf 86}, 044014 (2012), arXiv: 1207.6004;
 G.~J.~Olmo and D.~Rubiera-Garcia,
  Eur.\ Phys.\ J.\ C {\bf 72}, 2098 (2012), arXiv: 1112.0475;
 G.~J.~Olmo and D.~Rubiera-Garcia,
  Int.\ J.\ Mod.\ Phys.\ D {\bf 21}, 1250067 (2012), arXiv: 1207.4303;
 G.~J.~Olmo and D.~Rubiera-Garcia,
  Phys.\ Rev.\ D {\bf 88}, 084030 (2013), arXiv: 1306.4210;
 F.~S.~N.~Lobo, J.~Martinez-Asencio, G.~J.~Olmo and D.~Rubiera-Garcia,
  Phys.\ Lett.\ B {\bf 731}, 163 (2014), arXiv: 1311.5712.
 F.~S.~N.~Lobo, J.~Martinez-Asencio, G.~J.~Olmo and D.~Rubiera-Garcia,
  Phys.\ Rev.\ D {\bf 90}, 024033 (2014), arXiv: 1403.0105.

\bibitem{Bis} T. Biswas, E. Gerwick, T. Koivisto, and A. Mazumdar, Phys. Rev. Lett. {\bf 108}, 031101 (2012).
\bibitem{Bar}  C. Barragan and G. J. Olmo, Phys. Rev. D {\bf 82}, 084015 (2010); C. Barragan, G. J. Olmo, and H. Sanchis-Alepuz, Phys. Rev. D {\bf 80}, 024016 (2009).

\bibitem{frq} S.~M.~Carroll, A.~De Felice, V.~Duvvuri, D.~A.~Easson, M.~Trodden and M.~S.~Turner,
  Phys. Rev. D71, 063513 (2005)
  astro-ph/0410031;
G. Cognola, S. Zerbini, J. Phys. A45, 374014 (2012), arXiv: 1203.5032.

\bibitem{Clifton} T. Clifton and J. D. Barrow, 
Phys. Rev. D 72, 123003 (2005), gr-qc/0511076.

\bibitem{JSantos} M. Reboucas and J. Santos, 
Phys.Rev. D80, 063009 (2009), arXiv: 0906.5354.


\bibitem{Accioly} A. J. Accioly and A. T. Goncalves, 
J. Math. Phys. 28, 1547 (1987).

\bibitem{BarrowUn} J.~D.~Barrow, R.~Juszkiewicz and D.~H.~Sonoda,
  Mon.\ Not.\ Roy.\ Astron.\ Soc.\  {\bf 213}, 917 (1985).
	
\bibitem{Tasson} 	J.~D.~Tasson,
  Rept.\ Prog.\ Phys.\  {\bf 77}, 062901 (2014)
  [arXiv:1403.7785 [hep-ph]].

\end{thebibliography}
\end{document}